\documentclass[amsmath,amssymb,amsfonts,aps,pre,preprint,superscriptaddress,bibnotes,showpacs,showkeys,longbibliography]{revtex4-1}

\usepackage{graphicx}
\usepackage{subfigure}
\usepackage{amsmath}
\usepackage{amsfonts}
\usepackage{amssymb}
\usepackage{natbib}
\usepackage{relsize}
\usepackage{sidecap}
\usepackage{braket}
\usepackage{footmisc}
\usepackage{footnote}
\usepackage{color}
\begin{document}

\title{The forces on a single interacting Bose-Einstein condensate}

\author{Nguyen Van Thu}
\affiliation{Institute for Research and Development, Duy Tan University, Da Nang, Vietnam}
\affiliation{Department of Physics, Hanoi Pedagogical University 2, Hanoi, Vietnam}

\begin{abstract}
Using double parabola approximation for a single Bose-Einstein condensate confined between double slabs we proved that in grand canonical ensemble (GCE) the ground state with Robin boundary condition (BC) is favored, whereas in canonical ensemble (CE) our system undergoes from ground state with Robin BC to the one with Dirichlet BC in small-$L$ region and vice versa for large-$L$ region and phase transition in space of the ground state is the first order. The surface tension force and Casimir force are also considered in both CE and GCE in detail.
\end{abstract}

\maketitle

\section{Introduction\label{sec:1}}

The original Casimir effect is discovered by H. B. G. Casimir \cite{Casimir}, which caused by the confinement of vacumm fluctuations of the electromagnetic field between two slabs at zero temperature. In this case the author pointed out that Casimir force is attractive and varying as a power $\ell^{-4}$ with $\ell$ being inter-distance beetwen two slabs and it is proportional to area $A$ of the slab. A review for Casimir effect and its applications were mentioned in \cite{Bordag}.

In the field of Bose-Einstein condensate (BEC), there are many papers for this subject. For two component BECs, several interesting properties were studied in Ref. \cite{Thu1}. This work proved that Casimir force of a BECs is not simple superposition of the one of two single component BEC because of interaction between two species and especially this force vanished in limit of strong segregation.

For a single Bose-Einstein condensate, the Casimir effect has been considered in many aspects. Using field theory in one-loop approximation, Schiefele and Henkel \cite{Schiefele} expressed the Casimir energy as an integral of density of state, their result shown that this energy decays as $\ell^{-3}$. At finite temperature, this effect was also investigated \cite{Dantchev,Biswas2}. The Casimir force on an interacting Bose-Einstein condensate, which consists of mean field force and Casimir force, was calculated in Ref. \cite{Biswas3}, in which system under consideration was in grand canonical ensemble (GCE). However, as our understanding, the study on Casimir and mean field forces in canonical ensemble (CE) have been still absent so far. There are two main aims in this work: (i) consider surface tension force and Casimir force of a single BEC in both CE and GCE for both Dirichlet boundary condition (BC) and Robin BC and (ii) investigate the phase transition in space of the ground state within frame work of double parabola approximation (DPA). The system under consideration is a dilute interacting BEC \cite{Schiefele}.

This paper is organized as follow. In Section \ref{sec:2} we investigate the phase transition in space of the ground state, which depends on the applied BC. The forces act on the slabs, namely, surface tension force and force are studied in Section \ref{sec:3}. The conclusions and outlook are  given in Section \ref{sec:4} to close the paper.

\section{Boundary condition and phase transition in space of the ground state \label{sec:2}}

To begin with, we consider a single BEC confined between two parallel pallates of area $A$ along the $(x,y)$-plane and they are separated along $z$ direction by a distance $\ell$. For this geometry one requests $\sqrt{A}\gg \ell$. The positions of these slabs are $z=0$ and $z=\ell$. The total Hamiltonian reads
\begin{eqnarray}
{\cal H}=\int_V {\cal H}_b+\int_{W1}{\cal H}_{W1} dS+\int_{W2}{\cal H}_{W2} dS,\label{Hamilton}
\end{eqnarray}
in which ${\cal H}_b$ is Hamiltonian in bulk, without an external field, which has the form
\begin{eqnarray}
{\cal H}_b=\psi^*(\vec{r})\left[-\frac{\hbar^2}{2m}\nabla^2\right]\psi(\vec{r})+V_{GP},\label{Hb}
\end{eqnarray}
where
\begin{eqnarray}
V_{GP}=-\mu\psi(\vec{r})+\frac{g}{2}|\psi(\vec{r})|^4,\label{GPpotential}
\end{eqnarray}
is Gross-Pitaevskii (GP) potential.
Here we denote $\psi(\vec{r})$ is wave function of the ground state, which plays the role of order parameter, $m$ is atomic mass. The coupling constant $g$ is inter-particle interaction, which is determined via the $s$-wave scattering length $a_s$ through $g=4\pi\hbar^2a_s/m$. The chemical potential $\mu$ is read as $\mu=gn_0$ if our system contacts with "bulk reservoirs" of condensate, in other words, the system is considered in GCE. However in CE this chemical potential is determined by the relation for fixed particle number
\begin{eqnarray}
N=\int \psi^2 d\vec{r}.\label{normalize}
\end{eqnarray}
In mean field theory \cite{Schiefele,AoChui,Andersen}, this potential is derivative of free-energy density with respect to particle density and result is $\mu= gn_0$. Here we denote $n_0=N/V$ is bulk density and $V$ is volume of system.

${\cal H}_\alpha~(\alpha=W_1,W_2)$ are Hamiltonian of hard walls, which are chosen in the phenomenological forms \cite{Lipowsky, Binder},
\begin{eqnarray}
H_\alpha=\frac{\hbar^2}{2m\lambda_\alpha}\psi^*_\alpha\psi_\alpha,\label{Halpha}
\end{eqnarray}
with $\psi_\alpha$ being surface field at the slabs and and $\lambda_\alpha$ is extrapolation length.

Minimizing the total Hamiltonian (\ref{Hamilton}) leads to the time-independent Gross-Pitaevskii (GP) equation \cite{Pitaevskii},
\begin{eqnarray}
-\frac{\hbar^2}{2m}\nabla^2\psi(\vec{r})-\mu\psi(\vec{r})+g|\psi(\vec{r})|^3=0,\label{GP1}
\end{eqnarray}
and $\psi(\vec{r})$ fulfills boundary conditions at slabs \cite{Thuphatsong},
\begin{eqnarray}
\vec{n}\nabla\psi_\alpha=\frac{1}{\lambda_\alpha}\psi_\alpha.\label{BC}
\end{eqnarray}
The unit vector $\vec{n}$ perpendiculars to the surface at slabs and points inside the system.

It is worth noting that the condensate is translation along $(0x,0y)$-directions and the motion of particles are relevant only $z$-axis so that the nabla operator is replaced by derivative with respect to $z$, then Eq. (\ref{BC}) can be rewritten as follows
\begin{eqnarray}
\frac{\partial \psi_{W1}}{\partial z}\bigg|_{z=0}&=&\frac{1}{\lambda_{W1}}\psi_{W1}(z=0),\nonumber\\
\frac{\partial \psi_{W2}}{\partial z}\bigg|_{z=\ell}&=&\frac{1}{\lambda_{W2}}\psi_{W2}(z=\ell).\label{RBC}
\end{eqnarray}
These equations express the Robin BC at the slabs. When the surface fields are vanishing at slabs, which corresponds to Dirichlet ones.
\begin{eqnarray}
\psi(0)=\psi(\ell)=0.\label{DBC}
\end{eqnarray}
Eqs. (\ref{RBC}) and (\ref{DBC}) show that the BCs at slabs are either Robin BCs or Dirichlet BCs.

We now invoke the double parabola approximation (DPA) developed in \cite{Joseph} to study ground state of our system. To do this, we first introduce dimensionless coordinate $\varrho=z/\xi$ with $\xi=\hbar/\sqrt{2mgn_0}$ being healing length, the dimensionless order parameter $\phi=\psi/\sqrt{n_0}$. By this way, Eqs. (\ref{normalize}) and (\ref{GP1}) can be rewritten as
\begin{eqnarray}
&&-\partial_\varrho^2\phi-\phi+\phi^3=0,\label{GP2}\\
&&N=n_0\xi\int_0^L \phi^2d\varrho\equiv n_0\xi I_0,\label{normalize2}
\end{eqnarray}
where $L=\ell/\xi$. Next step, we note that near the slabs, because of decreasing from bulk value therefore we can expand the order parameter
\begin{equation}
\phi\approx 1+\delta,\label{khaitrien}
\end{equation}
with $\delta$ being a small real quantity. Putting (\ref{khaitrien}) into (\ref{GPpotential}) and keeping up to second order of $\phi$ one has DPA potential
\begin{eqnarray}
V_{DPA}=2(\phi-1)^2-\frac{1}{2}.\label{DPApotential}
\end{eqnarray}
At this step, instead of GP equation (\ref{GP1}) we have Euler-Lagrange equation
\begin{eqnarray}
-\frac{\partial^2\phi}{\partial\varrho^2}+\alpha^2(\phi-1)=0,\label{DPAGP}
\end{eqnarray}
where $\alpha=\sqrt{2}$.

Coming back to our problem, the system under consideration is symmetry, this means that $\lambda=\lambda_{W1}/\xi=-\lambda_{W2}/\xi$ and the Robin BCs (\ref{RBC}) are rewritten
\begin{eqnarray}
\phi(0)=\lambda\partial_\varrho\phi\bigg|_{\varrho=0},~\phi(L)=-\lambda\partial_\varrho\phi\bigg|_{\varrho=L}\label{RBC2}
\end{eqnarray}
in which $\lambda\geq0$. It is easily to find the solution for (\ref{DPAGP}) with constraint of (\ref{RBC2}), which is read as
\begin{eqnarray}
\phi=1-\frac{\cosh \left[(L-2\varrho)/\alpha\right]}{\alpha  \lambda  \sinh \left(\frac{L}{\alpha}\right)+\cosh \left(\frac{L}{\alpha}\right)}.\label{groundstate}
\end{eqnarray}

In order to determine $\lambda$ we require that when right slab goes to infinity the wave function (\ref{groundstate}) becomes exactly the one for semi-infinite system \cite{Thu}. Therefore one gets $\lambda=0$ and $\lambda=1/\alpha$ for Dirichlet and Robin BCs, respectively. A question raises naturally is that which one of BCs is in favor? The best answer will be given after calculating surface energy of the system. Firstly, we consider in GCE, the grand potential for the condensate is defined
\begin{eqnarray*}
\Omega=\int_{V}{\cal H}_bdV,
\end{eqnarray*}
hence
\begin{eqnarray}
\Omega=2P_0A\int_0^\ell \left[(\partial_\varrho\phi)^2+V_{DPA}\right]dz,\label{Omega1}
\end{eqnarray}
with $P_0=gn_0^2/2$ being the bulk pressure. Combining (\ref{DPApotential}) and (\ref{Omega1}) one has the excess energy (or surface tension) per unit area
\begin{eqnarray}
\gamma=\frac{\Omega-P_0V}{A}=2P_0\xi\int_0^L d\varrho\left[(\partial_\varrho\phi)^2+2(\phi-1)^2\right].\label{gammaGCE}
\end{eqnarray}
Substituting (\ref{groundstate}) into (\ref{gammaGCE}) we arrive
\begin{eqnarray}
\gamma=P_0\xi \frac{2 \alpha  \sinh (\alpha  L)}{\left[\alpha  \lambda  \sinh \left(\frac{L}{\alpha}\right)+\cosh \left(\frac{L}{\alpha}\right)\right]^2}.\label{gammaGCE1}
\end{eqnarray}
\begin{figure*}
  \mbox{
    \subfigure[\label{f1a}]{\includegraphics[scale=0.65]{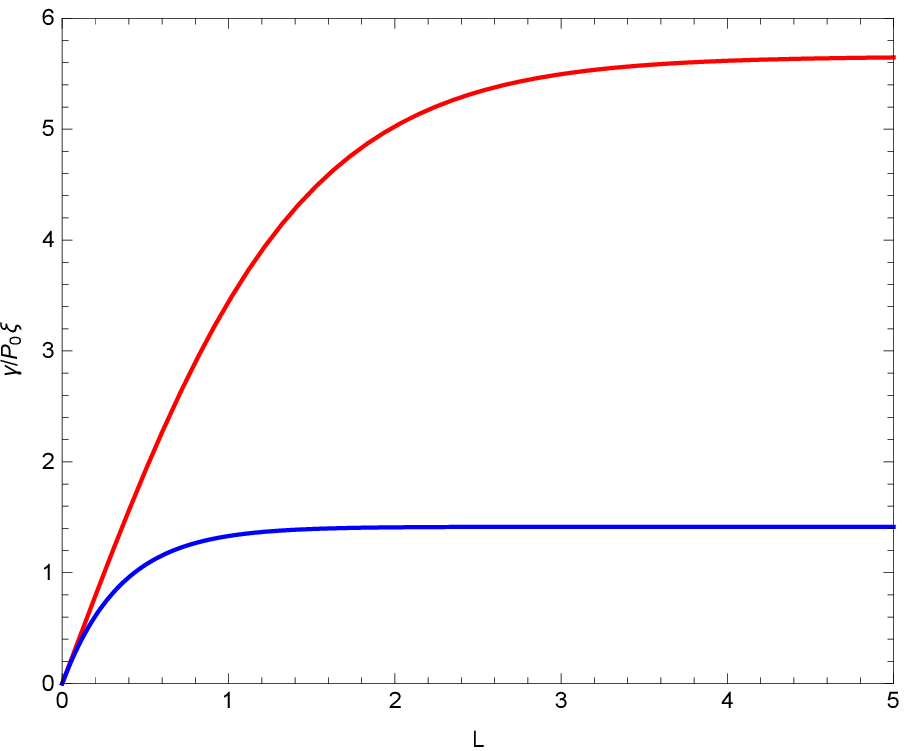}}\quad
    \subfigure[\label{f1b}]{\includegraphics[scale=0.65]{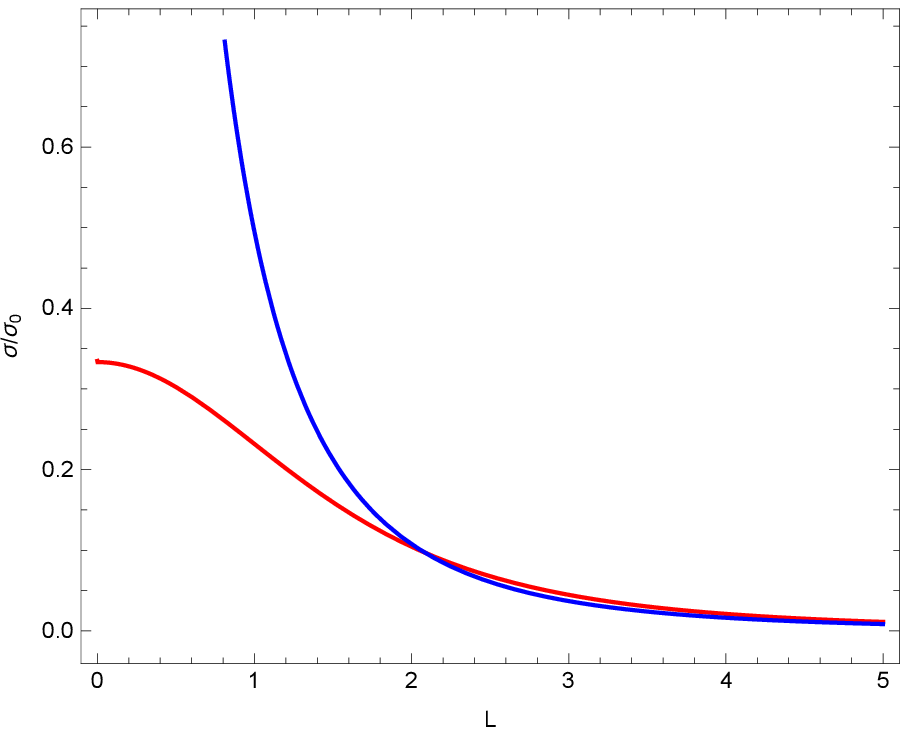}}
  }
  \caption{\footnotesize  (Color online) The surface tension of condensate in GCE (left) and CE (right). The red and blue lines correspond to $\lambda=0$ and $\lambda=1/\alpha$, respectively.}
  \label{f1}
\end{figure*}

\begin{figure*}
  \mbox{
    \subfigure[\label{f2a}]{\includegraphics[scale=0.65]{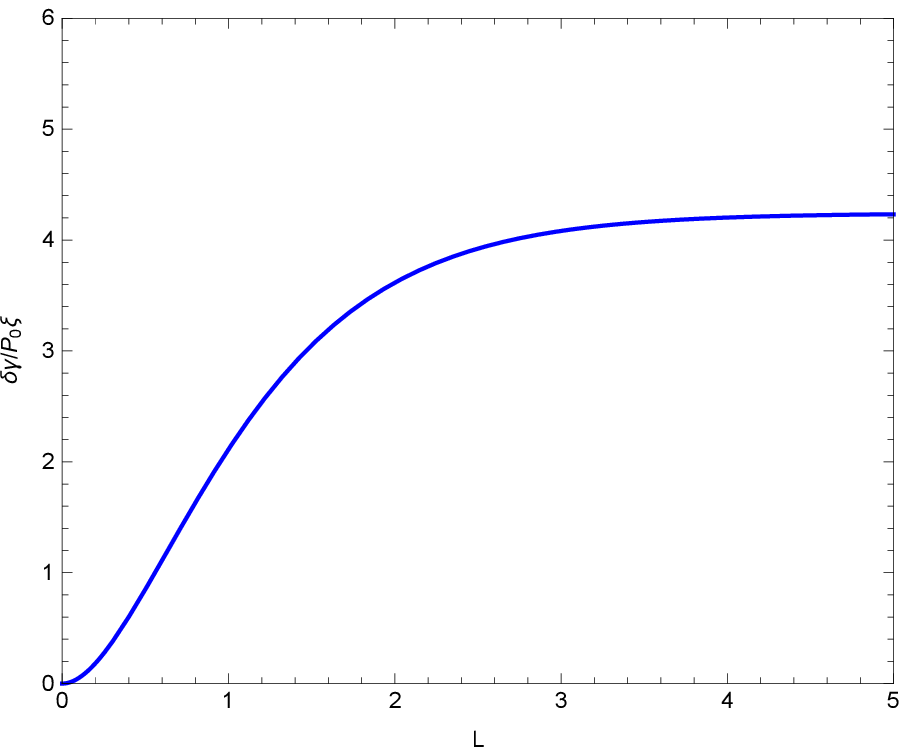}}\quad
    \subfigure[\label{f2b}]{\includegraphics[scale=0.65]{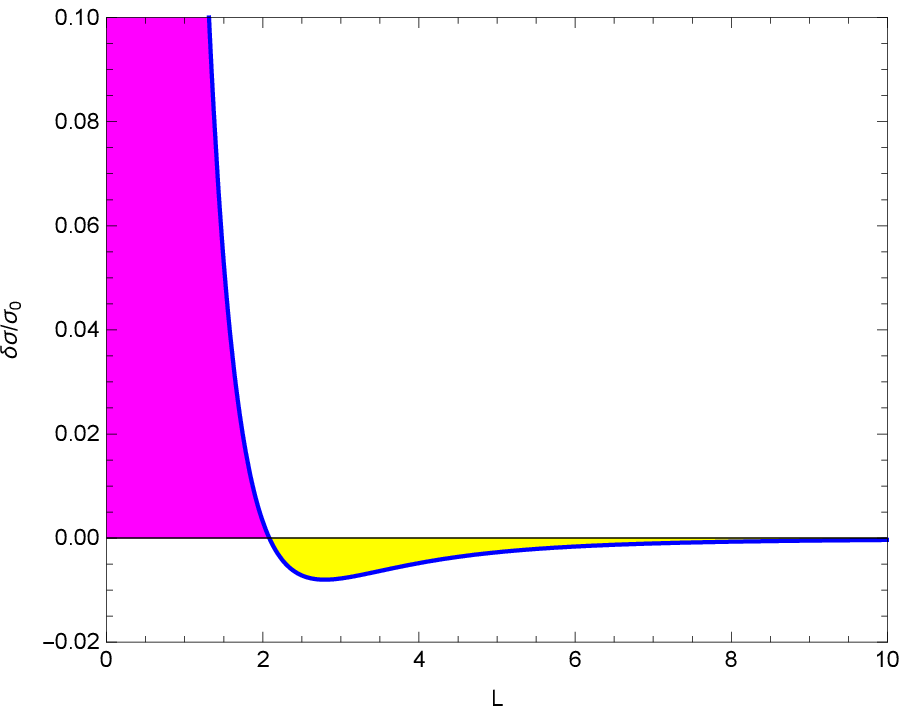}}
  }
  \caption{\footnotesize (Color online) The latent heat of phase transition in space of ground state for GCE (left) and CE (right).}
  \label{f2}
\end{figure*}

Fig. \ref{f1a} shows the $L$-dependence of surface tension for Dirichlet BC (red) and Robin BC (blue). At $L=0$ the surface tension is zero and it increases as $L$ increases and reaches constant when $L$ is large enough
\begin{eqnarray}
\lim_{L\rightarrow \infty}\gamma=\frac{4 \alpha }{(\alpha  \lambda +1)^2}P_0\xi.\label{gammaGCE2}
\end{eqnarray}
It is clearly that the surface tension corresponding to Robin BC is always smaller than the one corresponding to Dirichlet BC therefore the state corresponds to Robin BC is in favor. Impose that at the beginning our system is set up with Dirichlet BC, sooner or later it changes into the one with Robin BC and latent heat is
\begin{eqnarray}
\delta\gamma=\gamma[0,L]-\gamma[1/\alpha,L].\label{deltaEGCE}
\end{eqnarray}
The $L$-dependence of the latent heat is plotted in Fig. \ref{f2a}. Excepting for $L=0$, this latent heat differs from zero hence this phase transition is first order \cite{Thunew}.

We now focus on considering the surface tension in CE. A possible definition for surface excess energy was given by Ao and Chui \cite{AoChui}, by this means, the excess energy is total energy after a substraction of a contribution extensive in the volume
\begin{eqnarray}
 \Delta E=E_{CE}-\mu N=E_{CE}-N \frac{\partial E_{CE}}{\partial N}.\label{deltaE}
\end{eqnarray}
Combining Eqs. (\ref{deltaE}), (\ref{normalize2}) and (\ref{GP2}) then divided by area $A$ we obtain the surface tension
\begin{eqnarray*}
\sigma=\frac{\Delta E}{A}=\frac{1}{2}n_0\int_{0}^{\ell}\mathrm{d}z\phi\left(-\frac{\hbar^2}{2m}\nabla^2\right)\phi,
\end{eqnarray*}
or in dimensionless form
\begin{eqnarray}
\sigma=-P_0\xi\int_{0}^{L}\mathrm{d}\varrho \phi\partial_\varrho^2\phi.\label{tens2}
\end{eqnarray}
It is very interesting to distinguish this definition from the one which was proposed by Fetter and Walecka \cite{Fetter}. For the sake of simplicity, instead of (\ref{normalize2}), roughly speaking we impose that $n_0=N/A\ell$.  Inserting (\ref{groundstate}) into (\ref{tens2}) yielding
\begin{eqnarray}
\sigma=-\sigma_0 \frac{I}{L^3},\label{tensCE}
\end{eqnarray}
in which $\sigma_0=\frac{mg^2N^3}{\hbar^2A^3}$ and
\begin{eqnarray*}
I=\int_0^L \phi\partial_\varrho^2\phi d\varrho=\frac{\alpha  \left[\alpha  (2 \lambda +L)-2 \alpha  \lambda  \cosh (\alpha  L)-\sinh (\alpha  L)\right]}{2 \left[\alpha  \lambda  \sinh \left(\frac{L}{\alpha}\right)+\cosh \left(\frac{L}{\alpha}\right)\right]^2}.
\end{eqnarray*}
Fig. \ref{f1b} shows evolution of the surface tension versus $L$ in CE. The scenario is quite different in comparing with that in GCE. At $L=0$ the surface tension for Robin BC is divergent whereas it is finite for Dirichlet BC. In region $0<L<2.0834$ the surface tension for Robin BC is larger than that for Dirichlet BC and vice versa for remaining region. This means that in region $0<L<2.0834$ the ground state with Dirichlet BC is favor (magenta region in Fig. \ref{f2b}) and yellow region is supported for Robin BC. The latent heat for this phase transition is
\begin{eqnarray}
\delta\sigma=\sigma[0,L]-\sigma[1/\alpha,L],\label{deltaECE}
\end{eqnarray}
and it is sketched in Fig. \ref{f2b}.

\section{The force act on slabs\label{sec:3}}

In this section we consider the force acts on slabs, which consists of two components, namely, surface tension force caused by excess surface energy and Casimir force corresponding to the quantum fluctuation \cite{Biswas3}.

\subsection{Surface tension force}

The force corresponds to excess surface energy is defined as surface tension force. In GCE one has
\begin{eqnarray}
F_\gamma=-\frac{\partial\gamma}{\partial\ell}=-\frac{1}{\xi}\frac{\partial\gamma}{\partial L}.\label{forceGCE}
\end{eqnarray}
Plugging (\ref{gammaGCE1}) into (\ref{forceGCE}) we obtain
\begin{eqnarray}
F_\gamma=P_0 \frac{4\left[\alpha  \lambda  \sinh \left(\frac{L}{\alpha}\right)-\cosh \left(\frac{L}{\alpha}\right)\right]}{\left[\alpha  \lambda  \sinh \left(\frac{L}{\alpha}\right)+\cosh \left(\frac{L}{\alpha}\right)\right]^3}.\label{forceGCE1}
\end{eqnarray}
\begin{figure}[htp]
  \includegraphics{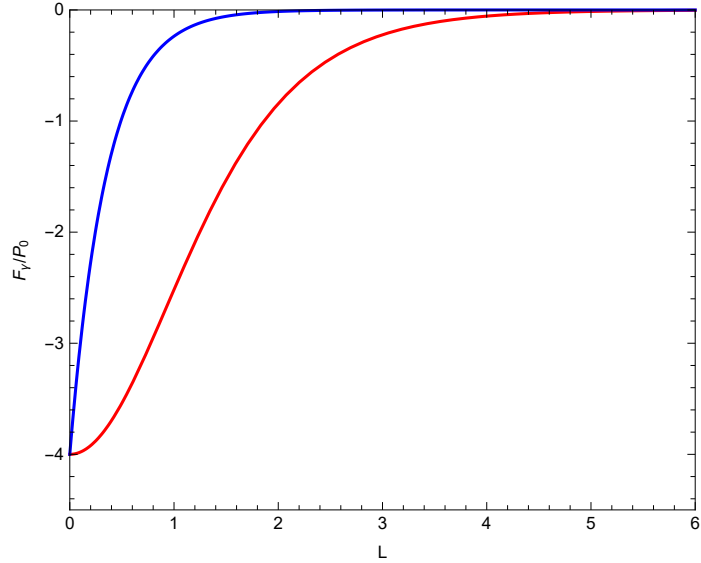}
  \caption{(Color online) The surface tension force in GCE versus $L$. The red and blue lines correspond to Dirichlet and Robin BC, respectively.}\label{f3}
\end{figure}
It is evident that the surface tension force generated by the leading order interaction term is given by $P_0=gn_0^2/2$ as mentioned in Ref. \cite{Pomeau}. The distance evolution of surface tension force are shown in Fig. \ref{f3} for Dirichlet BC (red line) and Ronin BC (blue line). It is obviously that these forces are always attractive. For all values of the distance $L$ surface tension force for Robin BC is smaller than that for Dirichlet BC, excepting for $L=0$ their values are the same and $F_\gamma=-4P_0$. When the distance increases there forces decrease sharply.

In CE, instead of $\gamma$, using $\sigma$ in (\ref{tens2}) one obtains
\begin{eqnarray}
F_\sigma=-\frac{1}{\xi}\frac{\partial\sigma}{\partial L}.\label{FCE1}
\end{eqnarray}
By this way, keeping (\ref{normalize2}) in mind, Eq. (\ref{FCE1}) can be rewritten as
\begin{eqnarray}
F_\sigma= F_0\frac{1}{L}\frac{\partial}{\partial L}\left(\frac{I}{L^3}\right),\label{FCE2}
\end{eqnarray}
in which $F_0=\frac{2m^2g^3N^4}{\hbar^4A^4}$.
\begin{figure}[htp]
  \includegraphics{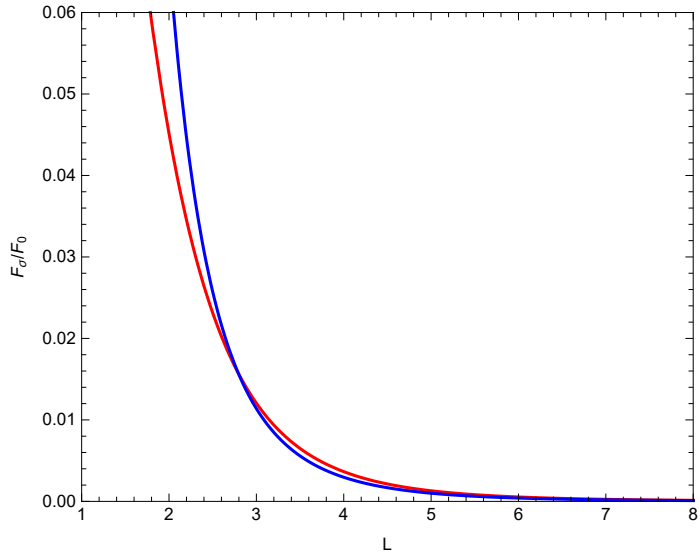}
  \caption{(Color online) The surface tension force in CE versus $L$. The red and blue lines correspond to Dirichlet and Robin BC, respectively.}\label{f4}
\end{figure}

Fig. \ref{f4} shows the surface tension force $F_\sigma/F_0$ as a function of distance $L$ at $\lambda=0$ (Dirichlet BC, red) and $\lambda=1/\alpha$ (Robin BC, blue). The situation basically differs from the one in GCE: surface tension force is repulsive and its strength decreases sharply as $L$ is creasing and the surface tension force for Robin BC is stronger than that for Dirichlet BC when distance $L$ is small enough. However, this force is divergent at $L=0$, the reason is the incompressibility of condensate.

\subsection{Casimir force}

We now consider the Casimir force caused by the quantum fluctuations on top of ground state, which corresponds to phononic excitations \cite{Schiefele,Biswas2,Biswas3,Biswas}. To do so, one employs the field theory in one-loop approximation, which has developed for the dilute single Bose gas \cite{Schiefele,Andersen} and two component Bose-Einstein condensates \cite{Thu1}. The Bogoliubov dispersion law for element excitation reads as
\begin{eqnarray*}
\varepsilon(\vec{k})=\sqrt{\frac{\hbar^2k^2}{2m}\left(\frac{\hbar^2k^2}{2m}+g\psi^2\right)},
\end{eqnarray*}
or, in dimensionless form
\begin{eqnarray}
\varepsilon(\kappa)=gn_0\sqrt{\kappa^2(\kappa^2+\phi^2)},\label{dispersion}
\end{eqnarray}
with dimensionless wave vector $\kappa=k\xi$. The density of free energy has the form
\begin{eqnarray}
\Omega=\frac{gn_{0}}{2\xi^3}\int\frac{d^3\vec{\kappa}}{(2\pi)^3}\sqrt{\kappa^2(\kappa^2+\phi^2)}.\label{term1}
\end{eqnarray}

Because of the confinement along $z$-axis, the wave vector is quantized as follows
\begin{eqnarray*}
k^2\rightarrow k_\perp^2+k_j^2,\label{k1}
\end{eqnarray*}
in which the wave vector component $k_\perp$ perpendicular to $0z$-axis and $k_j$ is parallel with $0z$-axis. In dimensionless form one has
\begin{eqnarray}
\kappa^2\rightarrow \kappa_\perp^2+\kappa_j^2.\label{k1}
\end{eqnarray}
It is well known that Gibbs and Helmholtz energies related to each other by a Legendre transform. Since we only consider here at zero temperature, i.e. only quantum fluctuation is taken into account hence the Casimir energy has the same form for both GCE and CE. Eq. (\ref{term1}) leads
\begin{eqnarray}
\Omega=\frac{gn_{0}}{2\xi^2}\sum_{j=-\infty}^\infty\int\frac{d^2\kappa_{\perp}}{(2\pi)^2}\sqrt{(\kappa_\perp^2+\kappa_j^2)(\kappa_\perp^2+\kappa_j^2+\phi^2)}.\label{term2}
\end{eqnarray}

In order to find the parallel component $k_j$ of wave vector, we consider the ideal Bose gas confined between two slabs. The wave function and energy correspond to eigenfunction and eigenvalue of Shrodinger equation
\begin{eqnarray*}
-\frac{\hbar^2}{2m}\frac{\partial^2 \Phi}{\partial z^2}=E\Phi.
\end{eqnarray*}
The wave function $\Phi$ is required to satisfy Robin BCs in (\ref{RBC}) with $\lambda_W\equiv\lambda_{W1}=-\lambda_{W2}$. It is easily to find that the wave vector has to be satisfied
\begin{eqnarray*}
k_j&=&\frac{\pi  j}{\lambda_W+\ell},
\end{eqnarray*}
or
\begin{eqnarray}
\kappa_j=\frac{\pi j}{\lambda+L}\equiv\frac{j}{\tilde{L}},~\widetilde{L}=\frac{\lambda+L}{\pi}.\label{k2}
\end{eqnarray}
Among other calculations \cite{Thu1}, Eq. (\ref{term2}) can be read
\begin{eqnarray}
\Omega=\frac{gn_{0}}{2\xi^2\widetilde{L}^2}\sum_{n=1}^\infty\int\frac{d^2\kappa_{\perp}}{(2\pi)^2}\sqrt{(\widetilde{L}^2\kappa_{\perp}^2+j^2)(M^2+j^2)},\label{term3}
\end{eqnarray}
where
\begin{eqnarray}
M=\widetilde{L}\sqrt{\kappa_{\perp}^2+\phi^2}.\label{massj}
\end{eqnarray}
Introducing a momentum cut-off $\Lambda$ for $\kappa_\perp$ we rewrite (\ref{term3}) in form
\begin{eqnarray}
\Omega=\frac{gn_0}{4\pi\xi^2\widetilde{L}^2}\int_0^\Lambda \kappa_\perp d\kappa_\perp\sum_{n=0}^\infty \sqrt{(\widetilde{L}^2\kappa_{\perp}^2+j^2)(M^2+j^2)}.\label{term4}
\end{eqnarray}
In order to calculate the Casimir energy (\ref{term4}), one employs the Euler-Maclaurin formula \cite{Pomeau} and takes a limit $\Lambda\rightarrow\infty$,
\begin{eqnarray}
\sum_{n=0}^\infty \theta_nF(n)-\int_0^\infty F(n)dn=-\frac{1}{12}F'(0)+\frac{1}{720}F'''(0)-\frac{1}{30240}F^{(5)}(0)+\cdots,
\end{eqnarray}
leads to
\begin{eqnarray}
\Omega=\frac{g n_0}{\xi ^2}\left[-\frac{\pi ^2 \phi }{1440 (\lambda +L)^3}+\frac{\pi ^4}{10080 \phi  (\lambda +L)^5}\right],\label{term5}
\end{eqnarray}
for both CE and GCE. Note that in (\ref{term5}), instead of $\widetilde{L}$ we used (\ref{k2}) to convert the result into $L$.

\begin{figure*}
  \mbox{
    \subfigure[\label{f5a}]{\includegraphics[scale=0.65]{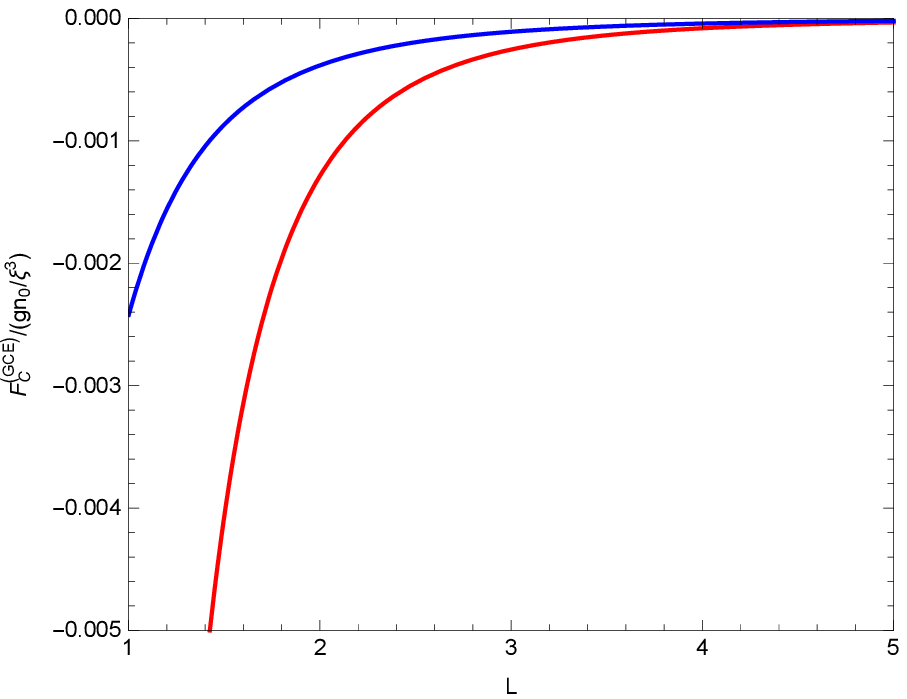}}\quad
    \subfigure[\label{f5b}]{\includegraphics[scale=0.65]{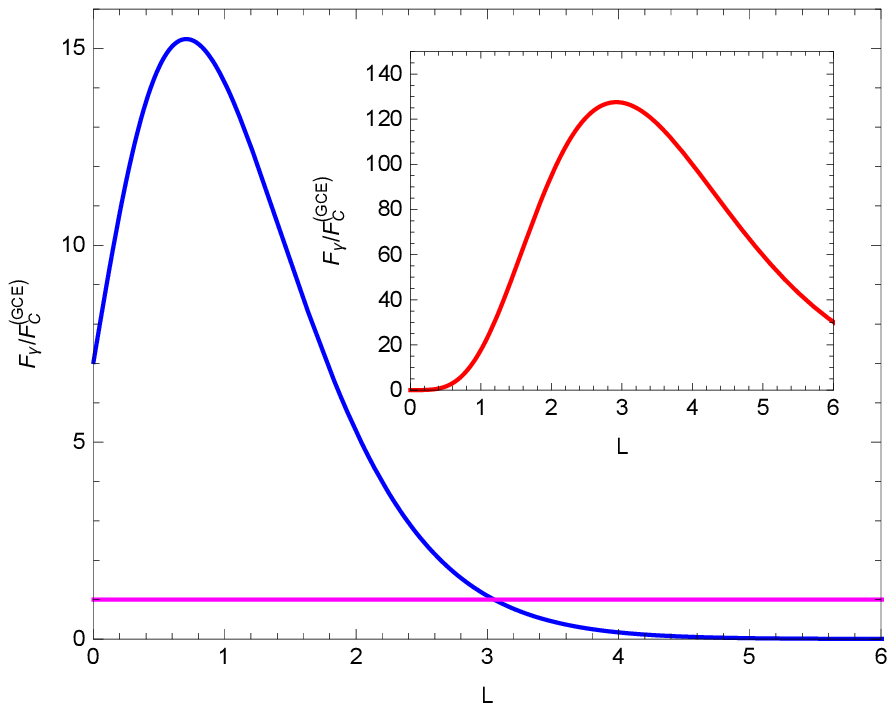}}
  }
  \caption{\footnotesize (Color online) The $L$-dependence of the Casimir force density in GCE (left) and ratio $F_\gamma/F_C^{(GCE)}$ (right). The red and blue lines correspond to Dirichlet and Robin BC, respectively.}
  \label{f5}
\end{figure*}

In GCE, because the bulk density of condensate is a constant, substituting (\ref{term5}) into (\ref{FCE1}) one obtains the density of Casimir force (per unit are of slab)
\begin{eqnarray}
F_C^{(GCE)}=\frac{g n_0}{\xi ^2}\left[-\frac{\pi ^2 \phi }{480 (\lambda +L)^4}+\frac{\pi ^4}{2016 \phi  (\lambda +L)^6}\right].\label{FC1}
\end{eqnarray}
Using the dimensional quantities Eq. (\ref{FC1}) gives
\begin{eqnarray*}
F_C^{(GCE)}\sim -\frac{\hbar v_s}{(\lambda_W+\ell)^4}+\frac{\hbar^2}{m^2v_s(\lambda_W+\ell)^6},
\end{eqnarray*}
in which $v_s=\sqrt{gn_0/m}$ is the speed of sound. For Dirichlet BC $\lambda_W=0$, this result coincides exactly with the one given in Ref. \cite{Pomeau}. Fig. \ref{f5a} shows the evolution of Casimir force density in GCE, where the red and blue lines correspond to Dirichlet and Robin BC.
\begin{figure}[htp]
  \includegraphics{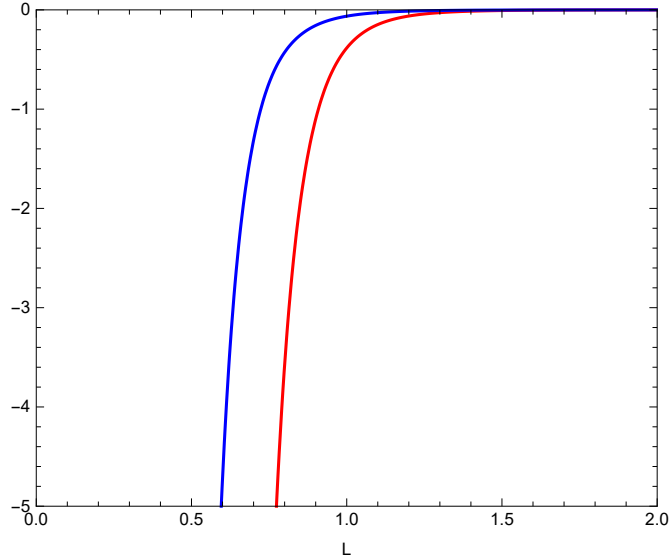}
  \caption{(Color online) The $L$-dependence of the Casimir force density in CE. The red and blue lines correspond to Dirichlet and Robin BC, respectively.}\label{f6}
\end{figure}
Based on these calculations and Fig. \ref{f5a} some remarkable comments are given:

- In Ref. \cite{Pomeau} the authors worked out the contribution due to the quantum fluctuation
\begin{eqnarray}
\Omega=\frac{1}{4\pi}\int_0^\Lambda \kappa_\perp d\kappa_\perp\sum_{n=0}^\infty\left\{\left[E_0^2(k)+2gn_0E_0(k)\right]^{1/2} -E_0(k)-gn_0\right\},\label{enPo}
\end{eqnarray}
in which $E_0=\hbar^2k^2/(2m)$ and $k^2=k_\perp^2+(j\pi/\ell)^2$. However, their result is the same as (\ref{FC1}), this means that two last terms in right hand side of (\ref{enPo}) have no contribution into the Casimir force.

- The Casimir force is attractive for both BCs and Casimir force for Robin BC is always smaller than the one for Dirichlet BC at a given value of the distance.

- When $L$ tends to zero, the density of Casimir force is divergent for Dirichlet BC whereas it is finite for Robin BC. At $L=0$ its value is
\begin{eqnarray}
F_C^{(GCE)}=-\frac{\pi ^2 g n_0 \phi }{480 \xi ^2 \lambda^4},\label{FC2}
\end{eqnarray}
when only the leading term in right hand side of (\ref{FC1}) is taken into account.

- In comparing to that in \cite{Biswas3}, in which the integral over was worked out with expanding the Casimir energy in power series of wave vector as shown in its Eq. (29) and keeping up to fourth order while the ultraviolet cutoff was taken to infinite limit, the first term in Eq. (\ref{FC1}) is agreeable but there is a difference in second term, instead of $L^{-5}$ our result gives $L^{-6}$.

- Combining Eqs. (\ref{forceGCE1}) and (\ref{FC1}) we see fraction of the Casimir force over the surface tension force in GCE is approximately
\begin{eqnarray*}
\frac{F_\gamma}{F_C^{(GCE)}}\sim \frac{\hbar^3}{2^{5/2}m^2gv_s}.
\end{eqnarray*}
Experimentally, consider for sodium \cite{Camacho} with $m=35.2\times 10^{-27}$ kg, $a_s=2.75\times 10^{-9}$ m one has $F_\gamma/F_C^{(GCE)}\approx 7.659$. The $L$-dependence of ratio $F_\gamma/F_C^{(GCE)}$ is sketched in Fig. \ref{f5b} for Robin BC, the insert shows that for Drichlet BC and magenta line corresponds to 1 in vertical axis. The parameters are chosen for sodium. It is obviously that, for Dirichlet BC the surface tension force is on top of the Casimir force, this coincides to comment in Ref. \cite{Biswas3}. For Robin BC one has: the surface tension force is stronger than Casimir force in region $L<3.053$ and vise versa for the other.

We now consider the Casimir force density in CE, in which only leading term in right hand side of (\ref{term5}) is kept. In this case, note that $n_0$ roughly depends on the distance $L$ via $n_0=N/A\ell$, therefore combining  (\ref{term5}) with the leading term and (\ref{FCE1}) one has
\begin{eqnarray}
F_C^{(CE)}=-\frac{m^2g^2N}{\hbar^4A}F_{0}\frac{\pi ^2 \phi  (4 \lambda +7 L)}{180 L^7 (\lambda +L)^4}.\label{FCE3}
\end{eqnarray}
The $L$-dependence of Casimir force density in CE is plotted in Fig. \ref{f6}, in which scaling for vertical axis is chosen $\frac{m^2g^2N}{\hbar^4A}F_{0}$. The red and blue lines correspond to Dirichlet and Robin BCs.
\begin{figure}[htp]
  \includegraphics{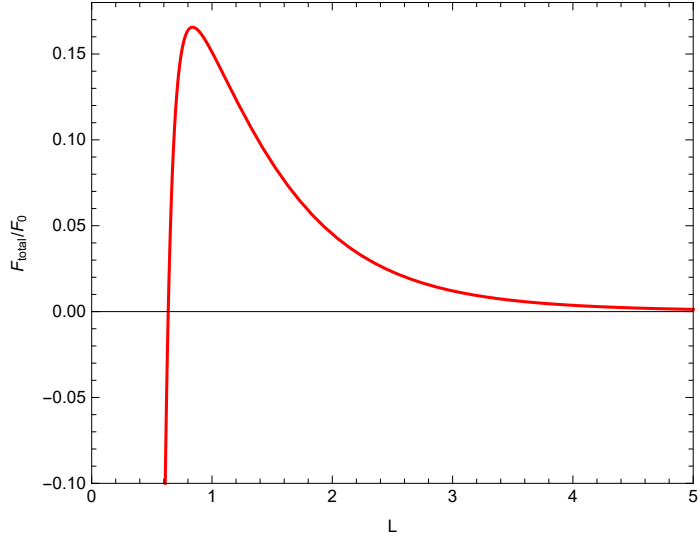}
  \caption{(Color online) The total force versus distance $L$ for Dirichlet BC in CE.}\label{f7}
\end{figure}

Although the Casimir force is attractive for both GCE and CE, there are several significant differences. Firstly, at $L=0$ this force is always divergent for both Dirichlet and Robin BCs in CE, whereas in GCE this force is finite if the Robin BC is applied. In addition, the Casimir force decays as increasing distance $L$ in law $L^{-4}$ and $L^{-9}$ for GCE and CE, respectively. Last but not least, at the same value of $L$, the Casimir force for GCE is much bigger than that for CE.

To end this section, let us compare the surface tension force with the Casimir force in CE. Combining (\ref{FCE2}) with (\ref{FCE3}) we first easily see that these forces are opposite, the surface tension force is repulsive and the other is attractive. The ratio of their strength approximately is
\begin{eqnarray*}
\frac{F_\sigma}{\left|F_C^{(CE)}\right|}\sim\frac{\hbar^4A}{m^2g^2N}.
\end{eqnarray*}
As already mentioned above, for sodium with total particle number $N=5.10^6$ and slab area $A=10^{-6} \text{m}^2$ one gets $F_\sigma/F_C^{(CE)}\sim 167.473$. It is worth noting that the surface tension force and Casimir one are opposite direction, furthermore both of them are divergent at $L=0$  hence one should calculate the total force $F_{total}=F_\sigma+F_C^{(CE)}$. As an example, we consider for Dirichlet BC and parameters of sodium, the graph is shown in Fig. \ref{f7}. This figure points out that the surface tension force is stronger than Casimir force in large-$L$ region and vice versa.

\section{Conclusion and outlook\label{sec:4}}

In the foregoing sections, using double parabola approximation and field theory we calculated the forces on slabs immersed in a single Bose-Einstein condensate. Our main results are in order

- The surface tension of the single Bose-Einstein condensate is obtained in both GCE and CE. The corresponding excess energy causes the surface tension force. This force is either attractive or repulsive, which depends on the system under consideration in GCE or CE.

- Based on the surface tension one finds that in GCE the ground state corresponds to Robin BC is favored whereas in CE the ground state is either Dirichlet BC for region small-$L$ or Robin BC for $L>2.0834$. The phase transition in space of the ground state is first order.

- The Casimir force is divergent at $L=0$ for both statistical ensembles. However this divergence disappears for Robin BC and in GCE.

One of our interesting result is that the phase transition in space of the ground state with no-zero latent heat, especially in CE, this transition produces exothermal or endothermal heat.

\section*{Acknowledgements}

I am grateful to Prof. Tran Huu Phat and Biswas for their useful discussions. This work is supported by Ministry of Education and Training of Vietnam.

\section*{References}

\bibliography{mybibfile}

\begin{thebibliography}{}
\bibitem{Casimir} H. B. G. Casimir,  Proc. K. Ned. Akad. Wet. {\bf 51}, 793 (1948).
\bibitem{Bordag} M. Bordag, U. Mohideen, V. M. Mostepanenko, Phys. Rep. {\bf 353}, 1 (2001).
\bibitem{Thu1} Nguyen Van Thu and Luong Thi Theu, J. Stat. Phys {\bf 168}, 1 (2017).
\bibitem{Schiefele} J. Schiefele, and C. Henkel, J. Phys. A {\bf 42}, 045401 (2009).
\bibitem{Dantchev} D. Dantchev, M. Krech, S. Dietrich, Phys. Rev. E {\bf 67}, 066120 (2003).
\bibitem{Biswas2} S. Biswas, Eur. Phys. J. D {\bf 42}, 109 (2007).
\bibitem{Biswas3} S. Biswas {\it et. al.}, J. Phys. B {\bf 43}, 085305 (2010).
\bibitem{AoChui} P. Ao and S. T. Chui, Phys. Rev. A {\bf 58}, 4836 (1998).
\bibitem{Andersen} J. O. Andersen, Rev. Mod. Phys. {\bf 76}, 599 (2004).
\bibitem{Lipowsky} R. Lipowsky, in Random Fluctuations and Pattern Growth, ed. by H. Stanley, N. Ostrowsky, NATO
ASI Series E, vol {\bf157} ( Kluwer Akad. Publ., Dordrecht, 1988), pp. 227–245
\bibitem{Binder} K. Binder, {\it in Phase Transitions and Critical Phenomena}, ed. by C. Domb, J. Lebowitz vol 8 (Academic
Press, London, 1983).
\bibitem{Pitaevskii} L. Pitaevskii and S. Stringari, {\it Bose-Einstein condensation}, Oxford University Press (2003).
\bibitem{Thuphatsong} N. V. Thu, T. H. Phat and P. T. Song, J. Low Temp. Phys. {\bf 186}, 127 (2017).
\bibitem{Joseph} J.O. Indekeu, C.-Y. Lin, N.V. Thu, B. Van Schaeybroeck, T.H. Phat, Phys. Rev. A {\bf 91}, 033615 (2015).
\bibitem{Thu} Nguyen Van Thu, Phys. Lett. A {\bf 380}, 2920 (2016).
\bibitem{Thunew} Nguyen Van Thu, Tran Huu Phat, HoangVan Quyet,{\it in preparation}.
\bibitem{Fetter} A. L. Fetter, J. D. Walecka, {\it Quantum theory of many-particle systems}, McGraw Hill, Boston 1971.
\bibitem{Pomeau} D. C. Roberts, Y. Pomeau, arxiv:cond-mat/0503757.
\bibitem{Biswas} S. Biswas, J. Phys. A {\bf 40}, 9969 (2007).
\bibitem{Camacho} A. Camacho, {\it Speed of sound in a Bose-Einstein condensate}, arxiv:1205.4774.
\end{thebibliography}

\end{document}